\documentclass[useAMS,usenatbib,usegraphicx]{mn2e}
\usepackage{epsfig}
\usepackage{amssymb}
\usepackage{amsmath}
\usepackage{times}

\begin{document}

\title[An Analysis Of Optical Pick--up In SCUBA Data]
{An Analysis Of Optical Pick--up In SCUBA Data}

\author[Zemcov, Halpern \& Pierpaoli]{
\parbox[t]{\textwidth}{
Michael Zemcov$^{1,2}$, 
Mark Halpern$^{1}$, 
Elena Pierpaoli$^{3}$
}
\vspace*{6pt}\\
$^{1}$ Department of Physics \& Astronomy, University of British
Columbia, Vancouver, BC V6T 1Z1, Canada \\
$^{2}$ School of Physics \& Astronomy, Cardiff University, Cardiff,
Wales CF24 3YB, UK \\
$^{3}$Physics Department and Astronomy Department, Princeton
University, NJ 08540, USA \\
\vspace*{0.cm}}

\date{Revised version \today}

\maketitle

\begin{abstract}
The Sub-millimetre Common User Bolometer Array (SCUBA) at the JCMT
employs a chopping and nodding observation technique to remove
variations in the atmospheric signal and improve the long term
stability of the instrument.  In order to understand systematic
effects in SCUBA data, we have analysed single--nod time streams from
across SCUBA's lifetime, and present an analysis of a ubiquitous
optical pick--up signal connected to the pointing of the secondary
mirror.  This pick--up is usually removed to a high level by
subtracting data from two nod positions and is therefore not obviously
present in most reduced SCUBA data.  However, if the nod cancellation
is not perfect this pick--up can swamp astronomical signals.  We
discuss various methods which have been suggested to account for this
type of imperfect cancellation, and also examine the impact of this
pick--up on observations with future bolometric cameras at the JCMT.
\end{abstract}


\section{Introduction}

In recent years, millimetre (mm) and sub-millimetre (sub-mm)
bolometric cameras have become the focus of much effort.  These
instruments directly detect thermal emission associated with star
formation and have proven useful in studying previously hidden
processes in our own galaxy and others to very high redshifts
(e.g.~\citealt{Lilly1999}, \citealt{Dunne2001}, \citealt{Johnstone2001},
\citealt{Ivison2002}, \citealt{Scott2002}, \citealt{Chapman2003}).
The Submillimetre Common User Bolometer Array (SCUBA;
\citealt{Holland1999}) at the James Clerk Maxwell Telescope (JCMT),
with its large aperture and excellent site, has been particularly
productive in this field.

Even exploiting a combination of high, dry sites and `windows' in
atmospheric opacity, emission due to the atmosphere and telescope's
surroundings is very large compared to astrophysical signals
\citep{Archibald2002}.  Moreover, it is highly variable on time scales
slower than a few Hz.  To counteract this, sub-mm telescopes can
employ a variety of observing strategies to scan the sky rapidly and
take differences to remove the bulk of the atmospheric emission.
These techniques put astrophysical signals into an audio frequency
band above the atmospheric 1/f knee.  Examples of specific observing
strategies are those employed by SHARC II \citep{Dowell2003}, which
scans the sky in a lissajous pattern, and BLAST \citep{Tucker2004} and
ACT \citep{Kosowsky2004}, which will scan the entire telescope in
azimuth only.  SCUBA uses a rapidly moving secondary mirror to chop
the beam rapidly and nods the whole telescope more slowly
\citep{Holland1999} while SCUBA2 (which will also be deployed at the
JCMT, \citealt{Holland2003}) intends to employ the DREAM algorithm
\citep{LePoole1998}.  In all of these approaches the observation and
data reduction processes which cancel atmospheric signals also filter
astrophysical signals at large angular scales.

\begin{figure}
\centering
\epsfig{file=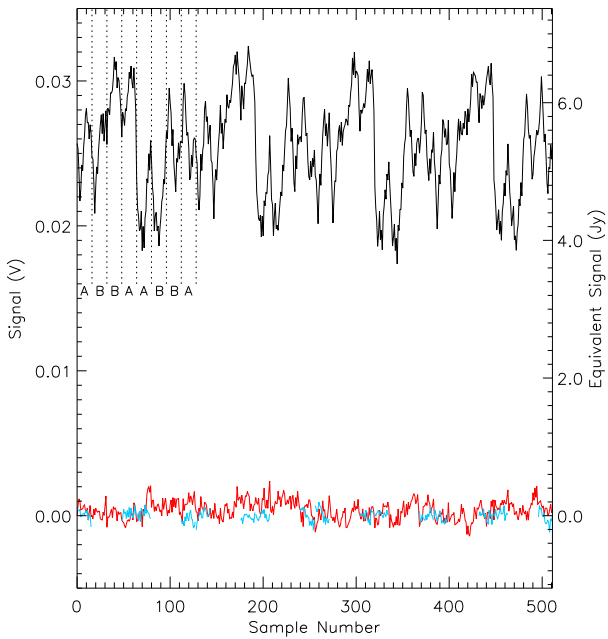,width=0.45\textwidth}
\caption{A typical reconstructed SCUBA data series for a $120$
arcsecond chop throw.  This particular data set is for bolometer G13
from observation 28 on 1998 June 30.  The upper trace shows the time
ordered data collected during four consecutive 64 point jiggle
patterns.  Each sample is $1.024 \,$s of data averaged over the chop
but there are occasional small gaps in the data series associated with
time spent nodding the telescope not shown here.  Every bolometer in
the array shows a similar pattern which is large, stable, and occurs
at the jiggle frequency.  In the usual data reduction pipeline
successive nod patterns are subtracted and this cancels the repetitive
pattern almost completely.  This is shown as the blue interrupted
trace which represents $A1 - B1$, $A2 - B2$, $A3 - B3$, and $A4 - B4$
for the four cycles.  There are therefore half as many data points in
this double--differenced set as in the raw data, so this trace is
plotted only in the $A$ nod position.  The lower red (continuous)
curve is the residual after subtracting the polynomial model of mirror
position discussed below.  It is nearly as clean as the blue data
set.}
\label{fig:timestream}
\end{figure}

Previous efforts to measure the Sunyaev-Zel'dovich (SZ) effect on the
JCMT have failed due to early removal of large angular scale signals
during data reduction.  Therefore, as part of an effort to measure the
SZ effect with SCUBA \citep{Zemcov2003}, we have taken the unorthodox
step of considering SCUBA data before subtracting data from adjacent
nods.  When we do this we find a repetitive signal which is orders of
magnitude larger than typical astronomical and even atmospheric
signals.  Fig.~\ref{fig:timestream} shows an example of this signal.
We will show that this signal is associated with changes in the
orientation of the secondary mirror and presumably arises from
altering the amount of thermal emission from the telescope which is
received by the bolometer array.  Although this signal is present in
all SCUBA data we have examined, it generally seems to be removed
effectively in the standard {\sc surf} \citep{Jenness1998} reduction
pipeline.  However, a period during which nod subtraction does not
effectively cancel this pick--up has already been identified by at
least one author \citep{Borys2004}.

This signal is synchronized with motion of the secondary mirror, and
therefore occurs in the data at a set of audio frequencies overlapping
those of astronomical signals.  Its presence imposes severe
constraints on viable observation strategies and on the required
precision of optical control at both the JCMT and similar telescopes.
In this paper we present measurements and analysis of this signal with
the aim of learning what these constraints are and determining the
origin of this signal.

\section{Observation of Optical Pick--Up}
\label{sec:data}

\subsection{SCUBA observation strategies}
\label{subsec:normal}

\citet{Holland1999} and \citet{Archibald2002} provide excellent
and extensive reviews of standard SCUBA operations and systematic
effects.  We review here just enough of the SCUBA observation modes to
allow understanding of the pattern shown in Fig.~\ref{fig:timestream}.

The JCMT's beam is chopped on the sky by nutating the secondary mirror
at about $7.8 \,$Hz in a nearly square wave pattern.  The angular
separation of the end points of this chop, called the \textit{chop
throw}, is chosen by the observer.  The chop is symmetric about the
optic axis of the telescope, which is to say that the two endpoints of
the chop pattern are each displaced from the optic axis by half the
chop throw.  The measured difference in intensity between the two
endpoints for every bolometer in both arrays is reported at
approximately $1 \,$Hz.  This signal is largely free of the
\textit{uniform} component of atmospheric emission.  We will call this
differential data the \textit{raw time series}.

The entire telescope is reoriented periodically to put one end of the
chop pattern and then the other on to the nominal source position.
This is called \textit{nodding} the telescope.  The difference in
intensity between the two nod positions forms a triple difference on
the sky: the brightness at the nominal source position minus that at
two symmetrically located off-positions.  This signal is largely free
of the effects of \textit{gradients} in atmospheric emission.  This
data set is called the \textit{de--nodded} or
\textit{double--differenced} data, and is also shown in
Fig.~\ref{fig:timestream}.

The SCUBA array under fills the focal plane, so the secondary mirror
undergoes a slow, small amplitude walk in addition to the movements
described above so that over time the sky is fully sampled.
\begin{figure}
\centering
\epsfig{file=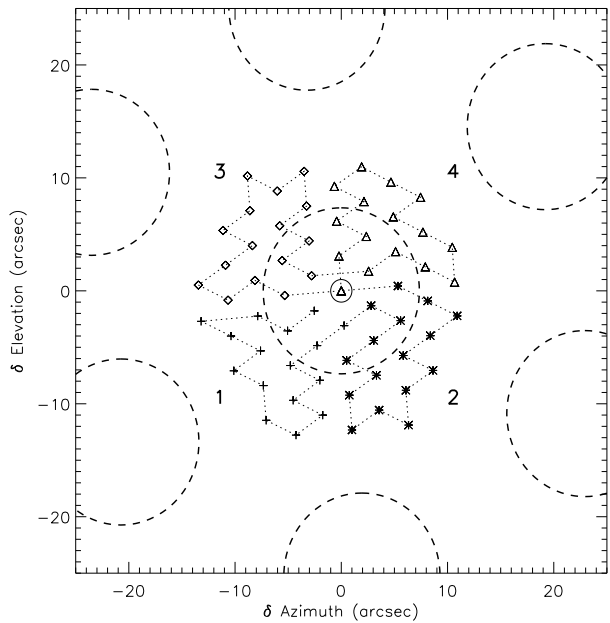,width=0.45\textwidth}
\caption{A schematic of the SCUBA jiggle pattern.  It is controlled by
motion of the secondary mirror, and allows the SCUBA array to fully
sample the sky.  The central dashed circle shows the full width half
maximum (FWHM) beam size associated with the (circled) central
triangular point.  The peripheral dashed circles show the FWHM of
adjacent bolometers in the SCUBA array; SCUBA under samples the sky at
any instant.  The secondary mirror chops the beam on the sky at $\sim
8 \,$Hz between positions separated by a user-defined angle, dwelling
for $\sim 1 \,$s with one end of the chop on each of the positions
shown above.  After completing the 16 positions in one quadrant of the
pattern the telescope primary mirror nods and the other end of the
chop is put onto each of the 16 positions in the same quadrant.  Then
the second, third, and fourth quadrants of the jiggle pattern are
observed in the same way.  If the primary mirror nod positions are
called $A$ and $B$ and the jiggle pattern quadrants are numbered 1
through 4 as in the figure, a 64 point measurement pattern consists of
the following sequence of $16 \,$s measurements: $A1 \, B1 \, B2 \, A2
\, A3 \, B3 \, B4 \, A4$.  This process requires approximately $128
\,$s to complete, plus some overhead associated with moving the
primary mirror.  Notice that the nod frequency of the primary mirror
is $\sim 1/32 \,$Hz while secondary mirror positions occur at
harmonics of $\sim 1/128 \,$Hz}
\label{fig:jigglepattern}
\end{figure}
Fig.~\ref{fig:jigglepattern} shows the standard 64 point pattern,
called a \textit{jiggle pattern}.  The 64 points are split into four
quadrants of 16 pointings each and the telescope nods after observing
each quadrant.  If we label these quadrants 1 through 4, and label the
two nod positions as $A$, and $B$, the order of data collection is
\begin{displaymath}
A1\quad B1\quad B2\quad A2\quad A3\quad B3\quad B4\quad A4\quad A1\dots
\end{displaymath}
The pattern repeats every $(16 \ \mathrm{Positions})\times(2 \ 
\mathrm{Nods})\times(4 \ \mathrm{Quadrants}) = 128$ samples, requiring
slightly longer than 128 seconds to complete one cycle.  The sample
series also has gaps in time because the primary mirror requires time
to move.  The order of observing nod $A$ and nod $B$ alternates by
quadrant to minimize primary mirror motion.  The peak to peak jiggle
distance is about 30 arcsec, and the average jiggle distance from the
middle of the pattern is about 9 arcsec. Very similar, albeit simpler,
motions occur in SCUBA's $16$ point jiggle pattern and 9 point
photometry modes.

Although the SCUBA bolometers are sampled at $128 \,$Hz, data are
reported as one second averages over the chop; undifferenced data are
not saved.  In standard operation, two time streams are reported, one
for each nod position.  Normally, the nod time streams are subtracted
from each other ($A1 - B1$ etc.), thereby removing a large component
of the atmospheric emission and instrumental variation.  This
double--difference time stream, which consists of the target beam data
minus half of each of the two chop positions' data, is the standard
SCUBA data product.

\subsection{The jiggle frequency signal}
\label{subsec:jiggle}

The raw time series in Fig.~\ref{fig:timestream} shows a clear
periodicity at 128 samples, indicating emission is modulated by the
128 point jiggle-and-nod pattern.  The peak to peak amplitude of the
pick--up signal is about $10 \,$mV, which is about $10^{4}$ times as
large as those arising from the weakest astronomical sources SCUBA is
capable of resolving.  In order to provide a consistent, if
approximate, guide by which to compare these signals to astrophysical
sources and thermal effects we have adopted a gain of $210 \,$Jy/V
throughout this paper.  This is consistent with the mean value
observed in pre-upgrade SCUBA data \citep{Jenness2002}.  With this
gain, the amplitude of the pick--up is about $2 \,$Jy per beam peak to
peak.

In the de--nodded data set there are 16 points in the set $A1 - B1$,
16 in the set $A2 - B2$, and so on.  These points are also plotted in
Fig.~\ref{fig:timestream}.  The amplitude of the pick--up signal has
been dramatically reduced and these signals are consistent with a
combination of detector and atmospheric noise.  We conclude that the
source of the pick--up is not modulated by nodding the telescope, and
therefore that the pick--up signal is associated with motions of the
secondary mirror only.

To check this conclusion, and to quantify the relation, we fit the raw
time series to a polynomial in a secondary mirror orientation using
elevation, $h$, and azimuth, $A$, coordinates as tracked by bolometer
H7's jiggle offsets, which are nominally zero on the optical axis of
the telescope.  This basis forms the natural coordinate system for
exploring a signal which arises within or around the telescope.  We
perform a fit independently in $h$ and $A$, using each angle and its
derivatives.  It is found that the fit does not require cross terms of
$A$ and $h$ to describe the data.  Our fitting function is therefore
\begin{multline}
\label{eqn:summands}
\tilde{V}(t) = V_{0} + \sum_{n=1}^{9} \alpha_{n} A^{n} +
\sum_{n=1}^{9} \beta_{n} \dot{A}^{n} + \sum_{n=1}^{9} \gamma_{n}
\ddot{A}^{n} + \\ \sum_{n=1}^{9} \delta_{n} h^{n} + \sum_{n=1}^{9}
\epsilon_{n} \dot{h}^{n} + \sum_{n=1}^{9} \zeta_{n} \ddot{h}^{n}
\end{multline}
where $(A(t), h(t))$ is the secondary mirror pointing at time $t$
measured in arcsec, dots denote time derivatives, and the constants
$\alpha_{n}$, $\beta_{n}$, $\gamma_{n}$, $\delta_{n}$, $\epsilon_{n}$
and $\zeta_{n}$ are determined from the fit.  In practice, these fits
are performed independently to $A$ and then $h$ for each bolometer.
The number of terms in the fit is arbitrary; we have chosen $n \leq 9$
as the ninth order terms are consistent with zero.  If the signal is
optical in origin, we expect the $n=1$ term, which corresponds to
position, to dominate, while if it is due to pick--up from motors or
electrical equipment, higher order terms, corresponding to velocities
and accelerations, should dominate.  Any pick--up associated with the
nod does not affect the time series at the jiggle frequencies, but
rather adds a variable DC offset between the two nod positions.  When
Eq.~\ref{eqn:summands} is fit, the linear terms $\alpha_{1}$ and
$\delta_{1}$ are at least an order of magnitude larger than any other
term.  This fit is therefore consistent with optical pick--up alone;
the higher order terms ($n > 1$) are included in order to demonstrate
that the pick--up pattern can be completely described in terms of the
position of the secondary mirror.  Further, we find that the pick--up
pattern is not constant but varies smoothly with bolometer position
(as shown below), which strongly suggests the pick--up is optical and
not electrical in origin.  Residuals from this fit, $V(t) -
\tilde{V}(t)$, are shown in Fig.~\ref{fig:timestream}.  Just like the
de--nodded data, these residuals also show nearly complete elimination
of the pick--up signal.  In a single data file this cancellation is
effective to about one part in 100; unfortunately, to observe weak
astrophysical signals in multiple integrations, we require
cancellation to better than one part in 3000.

\begin{figure}
\centering
\epsfig{file=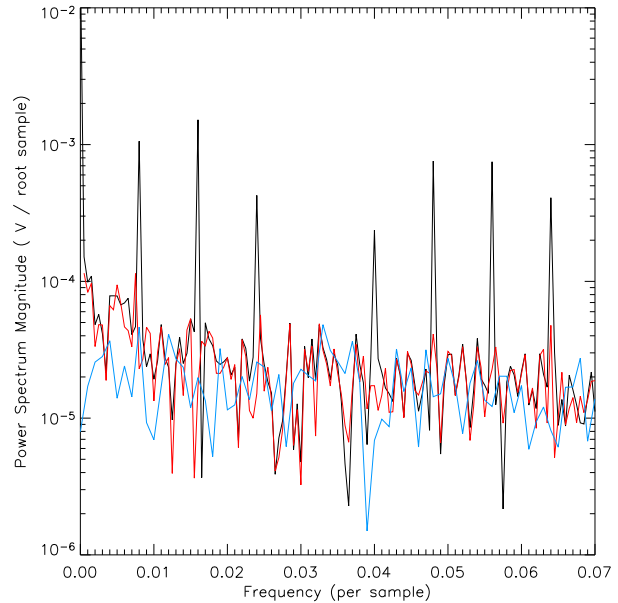,width=0.45\textwidth}
\caption{Power spectrum magnitudes (PSM, the modulus of the Fourier
transform) of the data shown in Fig.~\ref{fig:timestream}.  The
Fourier transforms have been taken with respect to sample number as
the precise mapping between sample number and time is complicated due
to small gaps in data collection during normal SCUBA operation.  As a
guide, 1 sample$\, \simeq 1 \,$Hz.  The black curve shows the PSM of
the raw data set.  Note that there is no power at $1/32$
sample$^{-1}$, confirming that the pick--up signal is associated with
secondary mirror position and not with motion of the primary mirror.
The red curve is the PSM of the residual signal after removal of the
polynomial model discussed below.  This curve contains no power in any
of the spikes in the raw data spectrum but faithfully reproduces the
PSM everywhere else.  The blue curve is the PSM of the double
differenced data.  Above $1/64$ sample$^{-1}$ the PSMs all have
comparable power, but at low frequency double differencing produces
`whitened' data with about half of the variance of the model
subtraction method.}
\label{fig:ffttime}
\end{figure}
Fourier transforms of the raw data, the de--nodded data, and the
residuals from Eq.~\ref{eqn:summands} are shown in
Fig.~\ref{fig:ffttime}.  Caution is required when interpreting the
transform of the de--nodded data since that data set is sampled
differently in time than the other two sets.  The raw data shows
spikes at harmonics of $1 / (128 \ \mathrm{samples})$.  Notice that
the fourth harmonic has no amplitude above the continuum level.  This
is because the transitions from $A1$ to $B1$, $A2$ to $B2$, etc.~occur
at the fourth harmonic and nodding does not alter the secondary mirror
pattern of motion.  All of the harmonic spikes are missing from the
power spectra of either the residuals or the de--nodded data.  Except
at the spikes, the fit residuals match the raw data with high
fidelity.  Notice that at very low frequencies the de--nodded data is
a bit quieter than the residuals are.  Presumably noise and large
angular scale astronomical signals which occur at low frequencies are
removed from the data by nod subtraction.

\section{Characterization of the pick--up Signal}
\label{sec:characterization}

In order to understand the severity and prevalence of this signal,
data were obtained from both the SCUBA archive\footnote{Courtesy the
Canadian Astronomy Data Centre, operated by the Dominion Astrophysical
Observatory for the National Research Council of Canada's Herzberg
Institute of Astrophysics.}, and a dedicated set of observations with
very small chop throw in 2003 June.  These data were chosen to reflect
a sampling of the types of mapping observation commonly performed with
SCUBA.  We have examined data taken in both 16 point and 64 point
jiggle patterns, as well as SCUBA's photometry mode which uses a small
9 point grid.  A variety of chop throw angles between 3 and 180 arcsec
have been chosen.  In order to reduce possible contamination by
astronomical sources, the data sets were selected based on low
estimated target flux.  An attempt was made to use only fields
containing sources with fluxes less than $10
\,$mJy, although this criterion was not met in the very small chop
throw (less than 15 arcsec) data.  A wide range of chop throws are
chosen to give a wide sampling of possible SCUBA jiggle map
configurations.  Also, data straddling the 1999 October SCUBA refit
and upgrade are considered.  All of the data were taken on photometric
nights, $0.04 < \tau_{\mathrm{CSO}} < 0.09$.  In all, over 200 data
files of varying lengths spanning nearly all of SCUBA's operational
lifetime have been considered in this analysis.

The data sets considered here have been preprocessed using {\sc
surf}\footnote{For a detailed description of {\sc surf}, see
\tt{http://www.starlink. rl.ac.uk/star/docs/sun216.htx/sun216.html}.}.
Programs are called to flat field, de-spike and extinction correct both
the double--differenced and single nod time streams; we direct the
reader to {\sc surf}'s documentation for the details of how these
procedures are implemented.  All of the data discussed here are left
at this stage of processing; advanced atmospheric removal has not been
implemented.

\subsection{Shape of the pattern}

\begin{figure*}

\centerline{
  \epsfig{file=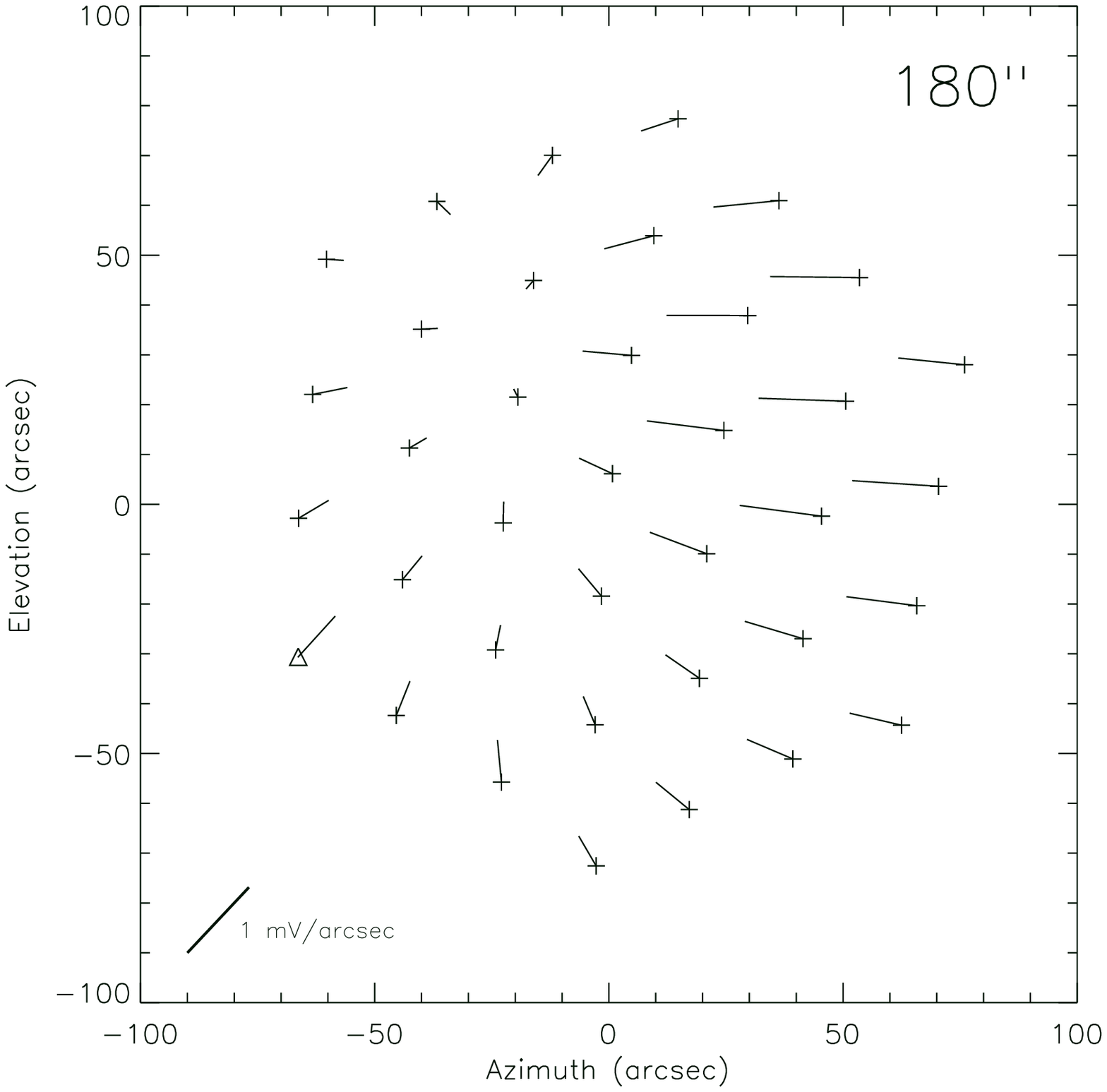,width=0.45\textwidth}
  \epsfig{file=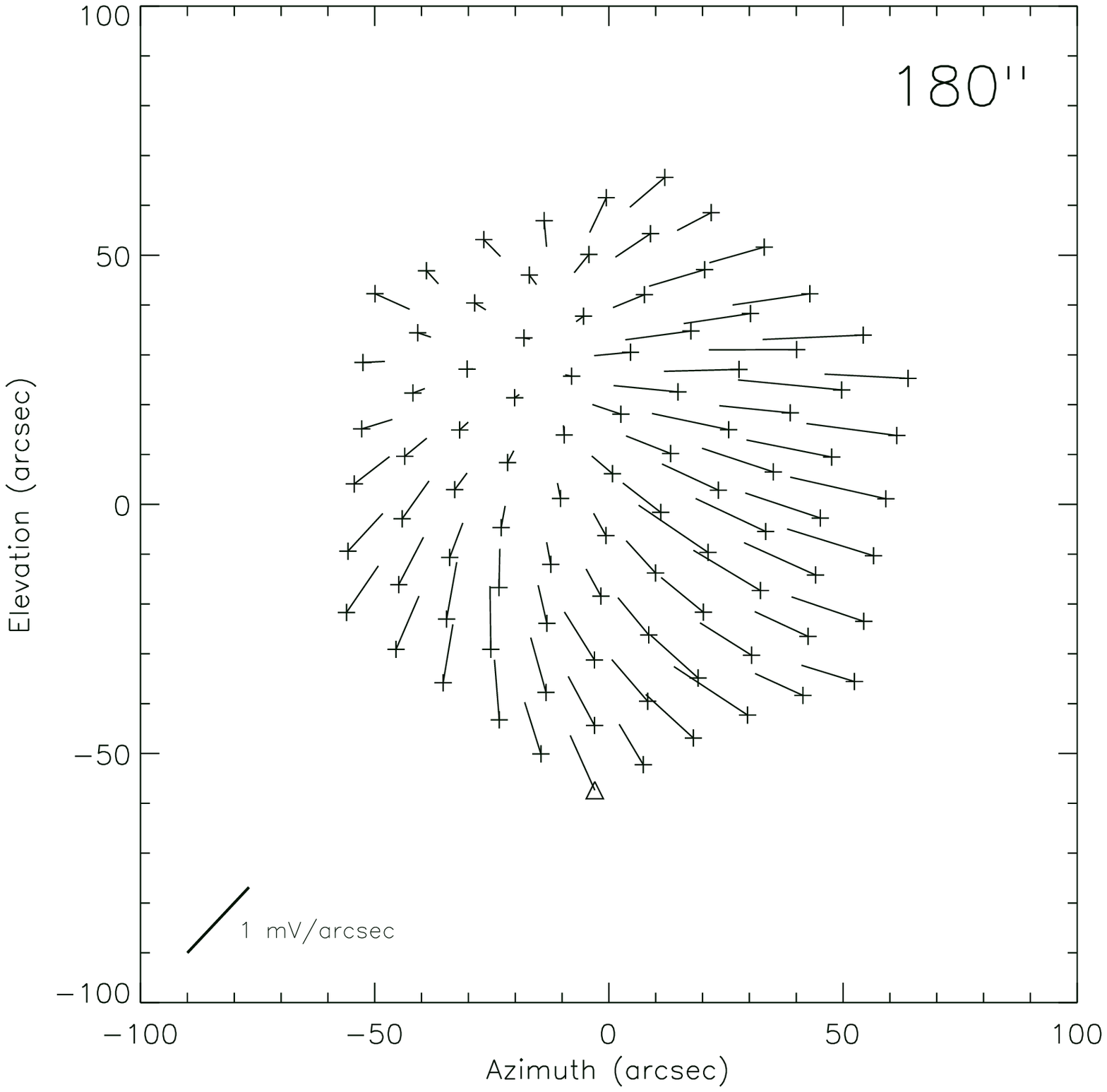,width=0.45\textwidth}
}
\centerline{
   \epsfig{file=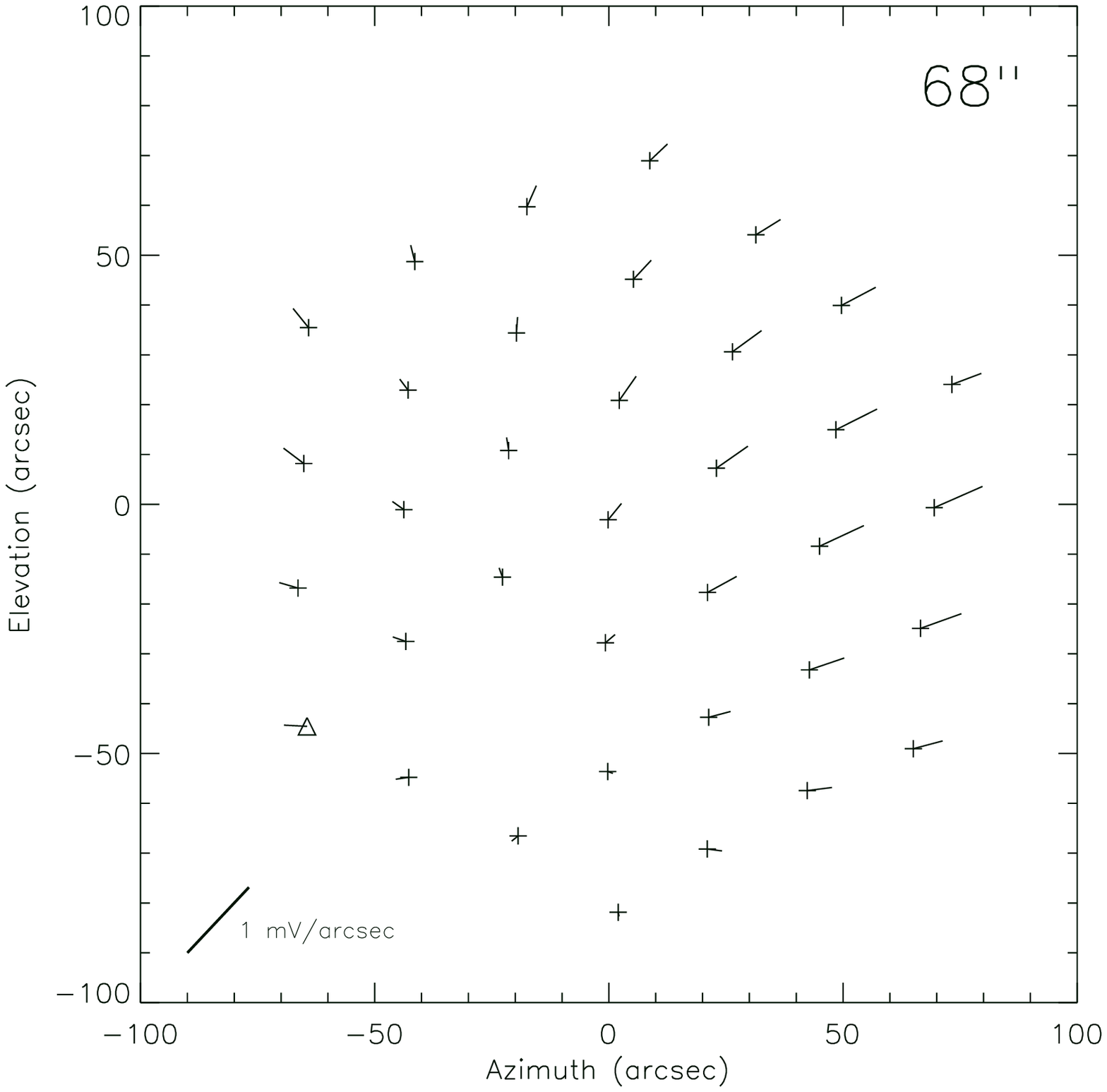,width=0.45\textwidth}
   \epsfig{file=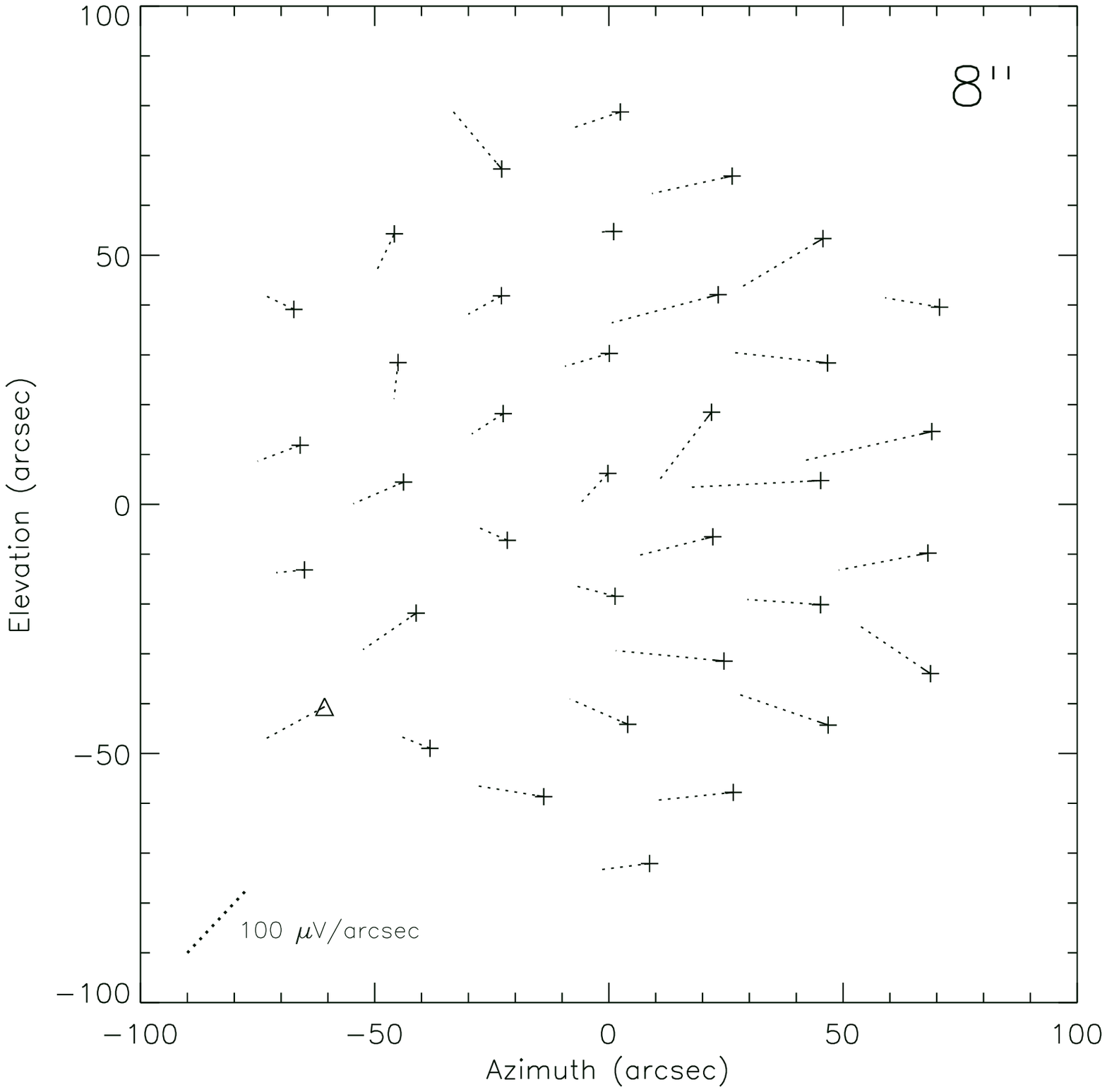,width=0.45\textwidth}
}

\caption{The set of $\mathbf{g}$ for a variety of chop throws (quoted
in the upper corner of each plot).  The top two plots are comparisons
of the $450$/$850 \, \mu$m arrays.  The plot on the upper left shows
the long wavelength array $\mathbf{g}$ from observation 67 on 1999
November 10, and the plot on the upper right shows its short
wavelength counterpart.  The $\mathbf{g}$ are shown beginning at each
bolometer location (`+' symbols).  Axis units apply only to the
bolometer locations, representative scales can be found in the lower
left of each plot.  Bolometer 1 is marked with a triangle symbol and a
scaling for $\mathbf{g}$ is shown in the bottom left of each plot for
reference.  The patterns are remarkably similar for both arrays,
suggesting that the jiggle frequency signal originates in SCUBA's
optical path rather after the dichroic beam-splitter
\citep{Holland1999} or in the electronics.  The bottom two plots are
for a 68 arcsec chop (observation 119 from 2003 June 14; left) and an
8 arcsec chop (observation 143 from 2003 June 14; right). In 2003,
bolometer G9 (number 7) was not operational; its data from that period
are not included.  These lower plots should be compared with the $180$
arcsec chop throw plot; on a gross scale, the patterns seem to be
similar over time.  However, smaller chops result in smaller pick--up
signals, and the exact pattern changes with each observation.}
\label{fig:patterns}
\end{figure*}

\begin{figure}
\centerline{
   \epsfig{file=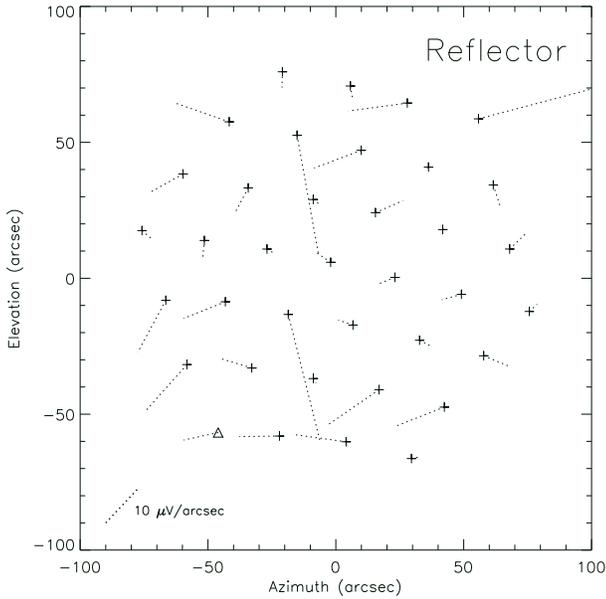,width=0.45\textwidth}
}
\caption{SCUBA pattern with a reflector blocking the dewar aperture
(observation 26 from 2002 October 10), with symbols as in Figure 3.
The $\mathbf{g}$ are consistent with being random, supporting the
claim that the jiggle pick--up originates outside of the SCUBA
cryostat.}
\label{fig:5}
\end{figure}

For each bolometer in a given observation, the vector 
\begin{eqnarray}
\mathbf{g} = \alpha_{1 \eta} \hat{A} +  \delta_{1 \eta} \hat{h},
\end{eqnarray}
where the subscript $\eta$ denotes the different bolometers, is
closely related to the gradient of $V$ in azimuth-elevation
coordinates.  Because the linear terms dominate, $\mathbf{g}$ captures
almost all of the power in $V$.  The units of $\mathbf{g}$ are V
arcsec$^{-1}$.  When the vector field $\mathbf{g}$ is plotted against
the location of each bolometer in the focal plane, a clear pattern
emerges (Figs.~\ref{fig:patterns} and \ref{fig:5}).  There is
typically a null in the patterns and the magnitude of $\mathbf{g}$
grows with separation from the location of the null in the focal
plane.

In order to test whether this signal is only present in the long
wavelength SCUBA array, the analysis discussed above is also performed
on the short wavelength array (Fig.~\ref{fig:patterns}).  It is
found that the patterns are remarkably similar between arrays, and in
particular the null in the pattern occurs at the same position in the
two focal planes during any given observation.  This implies that this
signal originates somewhere in SCUBA's optical path.

The pattern in Fig.~\ref{fig:patterns} is a derivative of the optical
loading on each bolometer.  There appears to be both a divergence and
a curl in the pattern, perhaps because Eq.~\ref{eqn:summands} does not
employ true partial derivatives, or because the secondary mirror's
orientation and a bolometer's location in the focal plane are not
identically related.  In any case, the location of the null in the
pattern corresponds to the location of an extremum in the optical
loading of the bolometer array.  It is striking that in our experience
this extremum does not occur at the array centre, which we presume is
the symmetry axis of the telescope.  However, since the telescope is
very difficult to align to arcsec accuracy\footnote{This is performed
either by using an alignment laser or, post-2002, by examining the
noise properties of the atmosphere.}, it is not unreasonable to expect
the symmetry axis to deviate from the array centre by a few tens of
arcsec (P.~Friberg, private communication).

As a routine part of our Sunyaev-Zel'dovich effect measurements, an
aluminum reflector was placed at the cryostat window and observations
with large chop throw were performed.  This eliminates all optical
inputs to the bolometer arrays but leaves electrical pick--up from the
various motors and acoustic pick--up intact.  The resulting pattern is
shown in Fig.~\ref{fig:5}.  The average amplitude of $\mathbf{g}$ is
more than two orders of magnitude smaller than without the reflector
in place.  The pick--up signal must be optical in origin.
%

The correlated $\mathbf{g}$ patterns occur in every SCUBA data set,
and gross features such as the magnitude of the vectors are constant.
However, details of each `vector field', such as the centroid, curl,
divergence, etc., vary slowly in time.  For instance, positional
changes in the vector field centroid of tens of arcsec can be tracked
over an evening of observation.  Furthermore, duplicate observations
of the same field performed months apart result in patterns which are
quite dissimilar.  We believe that this is because a given field's
altitude and azimuth have changed over the intervening time, even
though in sky-based coordinates the field's position is constant.  We
have had difficulty locating data taken with precisely the same
altitude, azimuth and chop throw separated by a suitable interval of
time, but surmise that such data would produce very similar patterns
if this is an optical pick--up related to ambient radiation in the
telescope dome and its surroundings.

\subsection{The $\mathbf{ \langle | g | \rangle }$ -- chop throw
distance relation}
\label{subsec:gtochop}

\begin{figure}
\centerline{
   \epsfig{file=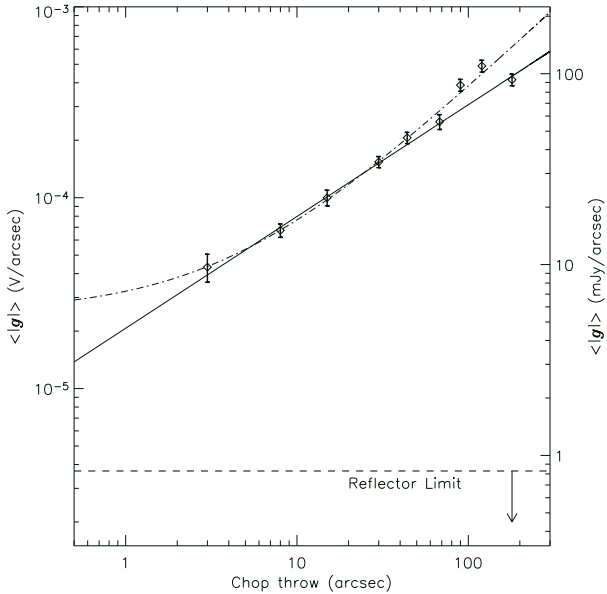,width=0.45\textwidth}
}
\caption{Jiggle frequency signal magnitude as a function of chop
throw.  The experimental points and their standard errors are shown as
diamonds with bars.  The lines show power law fits to the smallest
four chop throw data points.  The best--fitting power law with $\gamma = 0
\,$V arcsec$^{-1}$ offset is shown as the solid line.  Also, a power
power law with $\gamma = 2.5 \times 10^{-5} \,$V arcsec$^{-1}$
($\simeq 0.5 \,$mJy arcsec$^{-1}$) offset is shown as the dot--dash
line.  The dashed line and arrow at bottom indicate a lower limit to
this signal as determined from the reflector data.  Because the
reflector data's $\mathbf{g}$ are uncorrelated, $\langle | \mathbf{g}
| \rangle$ is an overestimate of the effect in this case, and we plot
the limit as half of that calculated.  Clearly, the jiggle frequency
signal has a substantial amplitude even at small chops.  It is very
difficult to ascertain the behavior of this signal as the chop throw
approaches 0 arcsec.}
\label{fig:opus}
\end{figure}

Careful analysis of Fig.~\ref{fig:patterns} shows that the mean
magnitude of the $\mathbf{g}$ vectors, $\langle | \mathbf{g} | \rangle
$, is correlated with chop size.  The quantity $\langle | \mathbf{g} |
\rangle $ and its standard deviation as a function of chop throw are
plotted in Fig.~\ref{fig:opus} for a sampling of data files with
different chop throws.


It is observed that the chop size and $\langle | \mathbf{g} | \rangle
$ are correlated.  While the characteristics of this correlation at
modest or large chops are relatively well defined, the signal's
behavior at very small chops is not so clear.  In an attempt to
clarify whether $\langle | \mathbf{g} | \rangle$ tends to zero at
small chop throws, power laws with an offset, $\gamma$
\begin{eqnarray}
\langle | \mathbf{g} | \rangle = \alpha \phi^{\beta} + \gamma.
\end{eqnarray}
are fit to data collected with chop throws $\phi$ less than 30
arcsec.  The parameters $\alpha$ and $\beta$ are determined from the
fit.  The results from two of these fits, one with an offset of $\gamma = 0
\,$V arcsec$^{-1}$ and the other with a voltage offset of $\gamma =
2.5 \times 10^{-5} \,$V arcsec$^{-1}$ (this value being arbitrarily
chosen for illustrative purposes) are listed in Table 1 and plotted in
Fig.~\ref{fig:opus}.  The $\chi^{2}$ statistics of these fits are
essentially the same.  Although the two functions differ in principle,
for chop throws of one arcsec the range allowed for the pick--up
amplitude is small, from $5$ to $7 \,$mJy per arcsec.  Even if the
pick--up pattern tends to zero at $\phi=0$, coherent mis--pointing
during a jiggle observation by a few per cent of the JCMT's FWHM
results in a spurious signal comparable to extra-galactic signals of
interest.

\begin{table}
\centering
\caption{Power law fit parameters.}
\begin{tabular}{ccccc}
\hline
Fit                         & $\gamma$          
& $\alpha$              & $\beta$         \\

                            & (V arcsec$^{-1}$) 
& (V arcsec$^{-1}$)     & (arcsec$^{-1}$) \\ \hline 

`No Offset'                 & 0.0               
& $2.07 \times 10^{-5}$ & $0.586$         \\

`$0.5 \,$mJy arcsec$^{-1}$' & $2.5 \times 10^{-5}$ 
& $7.35 \times 10^{-6}$ & $0.845$         \\ \hline

\end{tabular}
\end{table}

Interestingly, the power laws hold even for the large chop data,
although using data at large chop is a poor way to constrain the
characteristics of the signal at small chop (this is the motivation
for only using the smallest chop data in the fit).  While the presence
of an offset cannot be ruled out by these data alone, we suspect that
the relation is essentially linear and tends to zero at zero chop.

\section{Discussion}
\label{sec:Discussion}

Finding an optical pick--up signal which depends upon the orientation
of the secondary mirror is not a surprise.  Moving the secondary
mirror alters the electromagnetic cavity formed by the primary and
secondary mirrors of the telescope, and also alters the fraction of
the beam which intercepts the telescope structure or misses the
primary mirror.  As has been pointed out, the pattern in
Fig.~\ref{fig:patterns} is a derivative of total optical loading on
the bolometers.  It is quite striking that the observed pattern
sometimes corresponds to a minimum loading near to the optical axis,
and sometimes to a maximum (as in panels 1 and 3 of
Fig.~\ref{fig:patterns}).

To compare the observed signals to expectations based on a physical
understanding of the telescope first requires conversion of the
observations from Jy per beam to an equivalent Rayleigh--Jeans (RJ)
temperature.  The RJ brightness of a source at $850 \, \mu$m is $3.8
\times 10^{-17} \,$W m$^{-2}$ Hz$^{-1}$ str$^{-1}$ K$^{-1}$, and the
solid angle of the JCMT's primary beam is $\Omega = 1.13 \times
\mathrm{FWHM}^2 = 5.8 \times 10^{-9} \,$ str, so
\begin{multline}
\left( \frac{S}{\delta T} \right) = 5.8 \times 10^{-9} \ \mathrm{str}
\\ \times 3.8 \times 10^{-17} \, \mathrm{W \ m}^{-2} \ \mathrm{Hz}^{-1}
\ \mathrm{ K}^{-1} \ \mathrm{str}^{-1} = \\ 22.0 \ \mathrm{Jy \ K}^{-1}.
\end{multline}
Ideally, this conversion factor would be calculated as an integral over 
the SCUBA passband \citep{Borys1999}, which alters the numerical 
value by a few percent.

The $2.5 \,$Jy peak-to-peak signal in Fig.~\ref{fig:timestream}
corresponds to a $115 \,$mK RJ signal.  A more useful number for
understanding the emission mechanism is that the chop throw dependent
emissivity coefficient $\langle | \mathbf{g}| \rangle$ for a typical
chop throw of 60 arc seconds is
\begin{equation}
\langle | \mathbf{g}_{60^{\prime \prime}}| \rangle = 50 \, \mathrm{mJy
\ arcsec}^{-1} = 2.2 \, \mathrm{mK \ arcsec}^{-1}.
\end{equation}
Given that the telescope temperature is typically $270 \,$K, this
corresponds to a variation in emissivity of $10^{-5}$ per arcsec,
averaged over the array.

Several mechanisms could plausibly give emissivity variations which
are this high, and we do not know the SCUBA optical illumination
pattern well enough to make a definitive distinction.  It is certainly
possible that several processes are comparable in magnitude.  The
aluminum reflective surfaces are a few per cent emissive.  Their apparent
emissivity drops as the mirrors are tilted with respect to each other
giving rise to a maximum emission on the optical axis and a variation
which is to leading order quadratic in angle.  This mechanism will not
produce a minimum on axis.  However, the measured signal is the
difference between the gradients at two chop positions, and thus may
have either sign depending on the order of the difference.

The SCUBA plate scale is $\delta d = 1.5 \,$mm per arcsec (P.~Friberg,
private communication) so the illumination pattern on the primary
moves many cm during routine operation, altering the fraction of the
beam spilling off the edge of the mirror and terminating in the
telescope enclosure.  If $d B/d R$ is the fraction of the total SCUBA
response in an annulus of width $d R$ at the edge of the primary of
diameter $D = 15 \,$m, the anticipated signal from variations in beam
spill is
\begin{equation} 
\frac{dB}{B} = \frac{\pi D \delta d}{\pi D^2 / 4} \frac{dB}{dR} = 4
\times 10^{-4} \frac{dB}{dR} \, \mathrm{arcsec}^{-1}.
\end{equation}
This requires $dB/dR \approx 1$ per cent per mm, which is
\textit{possible} but larger than we expect.  

The secondary mirror support structure intercepts $8$ per cent of the
beam (P.~Friberg, private communication).  This coefficient could
easily vary by one part in $10^4$ per arcsec, which would produce a
signal at the observed level.  We do not know whether this effect
produces a maximum or a minimum near the optic axis.  In any case,
this very large optical signal is not a malfunction of the telescope,
but an unavoidable consequence of employing an observing strategy in
which the secondary mirror is moved.

\begin{figure}
\centerline{
   \epsfig{file=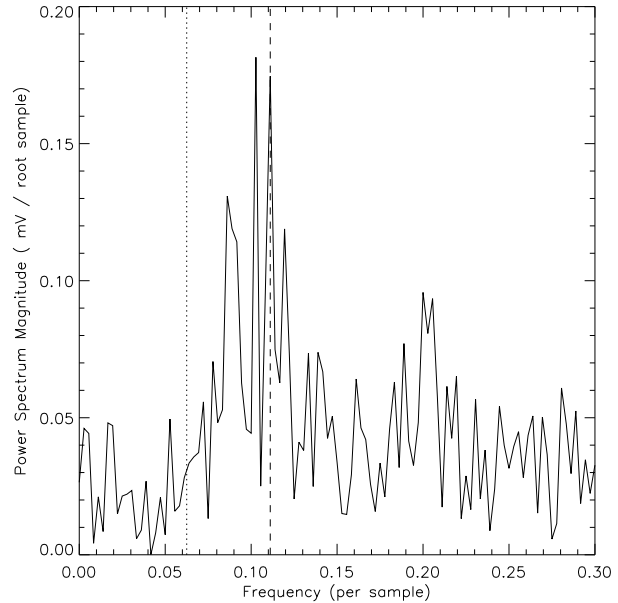,width=0.45\textwidth}
}
\caption{The PSM of typical reduced photometry data taken
between early to mid 2003 (bolometer I4 from run 57, 2003 February 9).
The jiggle frequency spike is present (shown as a dashed line at
$f=1/9$ sample$^{-1}$), as is a high level of broad-band noise between
about $0.05$ and $0.15$ sample$^{-1}$.  Both are caused by improper
nod cancellation; the optical pick--ups are brighter in one nod
position than the other, so the difference does not effectively remove
common mode contributions.  While the photometry mode jiggle frequency
is heavily contaminated, no spurious noise is present at the standard
jiggle map frequency (the dotted line at $f=1/16$ sample$^{-1}$).
This is further evidence that the pick--up is due to secondary mirror
position.}
\label{fig:phot}
\end{figure}

\citet{Borys2004} show that there is a spurious signal at a frequency near 
to 1 cycle per 16 samples present in SCUBA data collected during much
of 2003, and other groups agree (A.~Mortrier in preparation, T.~Webb
in preparation).  Published power spectra of this spurious signal look
strikingly similar to Fig.~\ref{fig:ffttime} given that the authors
have not re-ordered their data to be in time-order before implementing
a Fourier transformation. We presume that either hardware or software
difficulties during that period lead to imperfect nod cancellation,
leaving a substantial portion of the secondary mirror-induced pick up
signal present in the nominally cleaned data.  As a check, we have
examined photometry data collected during the same period.  In
Photometry mode, SCUBA executes a small nine-point grid to account for
small errors which might occur in the absolute pointing of the
telescope.  If the excess noise discussed in \citet{Borys2004} is due
to the secondary mirror and poor cancellation, there will be noise at
one cycle per nine samples in photometry data.  If the noise has a
different origin it might well be at the same audio frequency in
photometry mode as in jiggle mapping, i.e.~one cycle per 16 samples.
Fig.~\ref{fig:phot} shows the power spectrum of a typical bolometer in
the first set of photometry data from 2003 we have chosen to examine,
and the excess power is clearly near to the frequency of secondary
mirror motion for photometry.  This broad band of excess noise is
absent in data from either 1998 or 2000 which we examined.

We conclude that the spurious signal from 2003 is in fact poor
cancellation of secondary motion induced pick--up due to poor fine
control of the secondary mirror's postion.  This produces large
residual signals in the double--differenced data streams.  Because the
amplitude of the pick--up varies smoothly across the array
(e.g.~Fig.~\ref{fig:patterns}), we have investigated the possibility of
using the pattern as a template for removing the signal directly from
the data streams.  Unfortunately, no bolometer data seem to exhibit a
pick--up signal completely described by a gradient term, and it
appears that it is not possible to construct a time series that is
purely due to the pick--up from this template.

On a positive note, because this signal is bright and stable, it may
be possible to use it to provide a short term gain monitor of very
high signal to noise ratio in much the same way that
\citet{Jarosik2003} use the total power response of their radiometers
to obtain a short term gain monitor for \textit{WMAP} from the system
temperature.  However, a study of the stability of this signal and the
reliability of its dependence on operating parameters is required to
implement this strategy.

We do not know how large this effect is 4 arcmin off the optic axis
where SCUBA2 intends to have some sensitivity, but we presume from
Fig.~\ref{fig:opus} that it is very large. Furthermore, it is unclear
how well this signal can be disentangled from the atmospheric signal
without a chopper present in the system.  At this point, moving the
secondary in any way seems to require nodding or some other symmetric
differencing technique to remove the pick--up signal.  Unfortunately,
the DREAM algorithm currently under consideration for use with SCUBA2
at the JCMT could well be affected by this signal.  Furthermore,
instruments like THUMPER \citep{Walker2002} and BOLOCAM
\citep{Mauskopf2002}, which may be deployed at the JCMT, will have to
develop observation strategies which account for this type of
pick--up.  These are challenging issues which must be addressed before
the next generation of instruments is deployed at the JCMT.

\section*{Acknowledgments}

The team who commissioned SCUBA and developed the observing modes and
data reduction procedures deserve enormous credit for handling this
large systematic effect in a way that so nearly removes if from the
data.

Many thanks to Per Friberg, Colin Borys, Wayne Holland, Tim Jenness,
Walter Gear and Craig Walther for helpful discussions and advice. This
work was funded by the National Sciences and Engineering Research
Council of Canada.  EP was supported by a CITA National Fellowship
over the course of this work.  MZ is a guest user of the Canadian
Astronomy Data Centre, which is operated by the Dominion Astrophysical
Observatory for the National Research Council of Canada's Herzberg
Institute of Astrophysics.  The James Clerk Maxwell Telescope is
operated by the Joint Astronomy Centre on behalf of the Particles
Physics and Astronomy Research Council of the United Kingdom, the
Netherlands Organization for Scientific Research, and the National
Research Council of Canada.  We would like to acknowledge the staff of
the JCMT for facilitating our observations, and K.~Coppin for
performing them with very little warning and even less sleep.

\parskip 10pt 
\noindent This paper has been typeset from a \TeX/\LaTeX file prepared
by the author. 

\end{document}